\pdfoutput=1
\documentclass[11pt]{article}

\usepackage[T1]{fontenc}
\usepackage[utf8]{inputenc}
\usepackage{lmodern}
\usepackage[a4paper,margin=1in]{geometry}
\usepackage{amsmath,amssymb,mathtools,bm}
\usepackage{booktabs,tabularx,array,multirow}
\usepackage{graphicx}
\usepackage{xcolor}
\usepackage{microtype}
\usepackage{enumitem}
\usepackage[numbers,sort&compress]{natbib}
\usepackage[colorlinks=true,allcolors=blue!55!black]{hyperref}
\hypersetup{
  pdftitle={Gauge couplings of the Standard Model in the octonionic framework: invariant normalisations, current-consistent abelian matching, and a scale-explicit conditional ledger},
  pdfauthor={Tejinder P. Singh},
  pdfsubject={Octonionic unification and Standard Model gauge couplings},
  pdfkeywords={octonions, E8, gauge couplings, fine-structure constant, strong coupling, weak mixing angle}
}
\usepackage[nameinlink,capitalise,noabbrev]{cleveref}
\usepackage{caption}
\usepackage{placeins}
\usepackage{float}

\graphicspath{{figures/}}
\setlist{nosep,leftmargin=*}
\newcommand{\OO}{\mathbb O}
\newcommand{\CC}{\mathbb C}
\newcommand{\RR}{\mathbb R}
\newcommand{\Tr}{\operatorname{Tr}}
\newcommand{\MSbar}{\overline{\mathrm{MS}}}
\newcommand{\alphastar}{\alpha_{\star}}
\newcommand{\gstar}{g_{\star}}
\newcommand{\EomE}{E_8\!\times\!E_8}

\title{Gauge couplings of the Standard Model in the octonionic framework:\\
invariant normalisations, current-consistent abelian matching,\\
and a scale-explicit conditional ledger}
\author{Tejinder P. Singh\\
\small Tata Institute of Fundamental Research, Homi Bhabha Road, Mumbai 400005, India\\
\small \texttt{tpsingh@tifr.res.in}}
\date{8 October 2026}

\begin{document}
\maketitle

\begin{abstract}
We present a normalisation audit of the gauge-sector account of the $\EomE$ octonionic unification programme.  Exact identities within the specified algebraic construction, consequences of an effective support model, and microscopic hypotheses are kept separate.  The recurrent $3/8$ is the squared half-spacing of the adopted exceptional-Jordan family spectrum, while the exponential charge dependence is a multiplicative-character hypothesis motivated by the trace-dynamics action; the action parameter $\alpha_{\rm TD}$ is not the electromagnetic fine-structure constant.  On the real ladder space $H_6$, the real Schur lemma fixes the \emph{shape} of a colour-singlet self-adjoint photon profile to be proportional to the identity.  The resulting $1/6$ dilution follows only if the Hilbert--Schmidt norm used in the support model is the physical norm induced by the projected gauge kinetic term.  It therefore gives the conditional boundary value $[\alpha_\gamma^{(0)}]^{-1}=137.04006064\ldots$, whereas comparison with CODATA requires the factor $Z_\gamma=1.0000296379\ldots$, which remains to be calculated independently.

We also repair the two-$U(1)$ matter-current normalisation.  For parent charges $q_1=c_1Y$, $q_2=c_2Y$ with $c_1+c_2=1$, a link field of charges $(+q_\ell,-q_\ell)$ leaves the diagonal connection coupled to exactly the Standard-Model generator $Y$.  Its coupling is fixed by the complete kinetic matrix.  Equal \emph{connection} coefficients and zero kinetic mixing give the illustrative benchmark $g_Y^2/g_2^2=3/10$ and $\sin^2\theta_W=3/13$; however, trace-induced normalisation of the scaled generators gives a different result.  Thus $3/13$ is not an embedding-independent prediction.  If treated as exact, it is statistically incompatible with the quoted $\MSbar$ value by about $13$ standard deviations.  The colour-profile bound is likewise restricted to the strengthened ansatz in which each individual colour profile commutes with the complex structure; adjoint covariance alone does not imply it.

Measured couplings are used as inputs for scale and running diagnostics, not as predictions.  The assignment of the electromagnetic boundary to $Q\to0$ and the colour boundary to $M_Z$ remains phenomenological.  A two-loop gauge-running cross-check and three-loop QCD evolution leave the qualitative conclusions unchanged, while electroweak threshold matching remains spectrum dependent.  Recent localisation and six-dimensional tests identify sharper missing calculations but do not generate either the photon kinetic norm or $Z_\gamma$.  The result is a falsifiable conditional programme and a normalisation audit, not a parameter-free derivation of Standard-Model gauge couplings.
\end{abstract}

\tableofcontents

\section{Purpose and status of the argument}
\label{sec:purpose}

The octonionic programme seeks to obtain the Standard Model and a right-handed pre-gravitational sector from a pre-quantum matrix dynamics (a trace dynamics in the sense of Adler~\cite{Adler2004,AdlerMillard1996}) on split-bioctonionic spinor space, with an unbroken $\EomE$ structure before localisation and symmetry breaking~\cite{Singh2022EPJP,RajSingh2022,Kaushik2022,SinghBook2026,SinghSpectral2026}.  The recent spectral-action construction supplies a conditional route from the generalised trace-dynamics sectors to the structural low-energy action, but explicitly assumes the localisation map and broken-phase support factors~\cite{SinghSpectral2026}.  The gauge-sector calculations of this programme contain three numerically interesting observations:
\begin{enumerate}
\item an exceptional-Jordan/exponential expression close to the low-energy fine-structure constant~\cite{Singh2022EPJP};
\item a broken-phase support construction giving the conditional ratio $\alpha_s/\alpha_{\rm em}=16$ (versions 1--3 of the present paper);
\item a spinorial weak-angle formula giving $\sin^2\theta_W\simeq0.2497$~\cite{RajSingh2022}.
\end{enumerate}
The first two can be reorganised into a coherent strong--electromagnetic boundary ledger, although important hypotheses remain.  The third requires replacement: the value is in pronounced tension with electroweak data, and the derivation is not invariant under field normalisation (\cref{sec:weak-nogo}).

\begin{figure}[t]
  \centering
  \includegraphics[width=\textwidth]{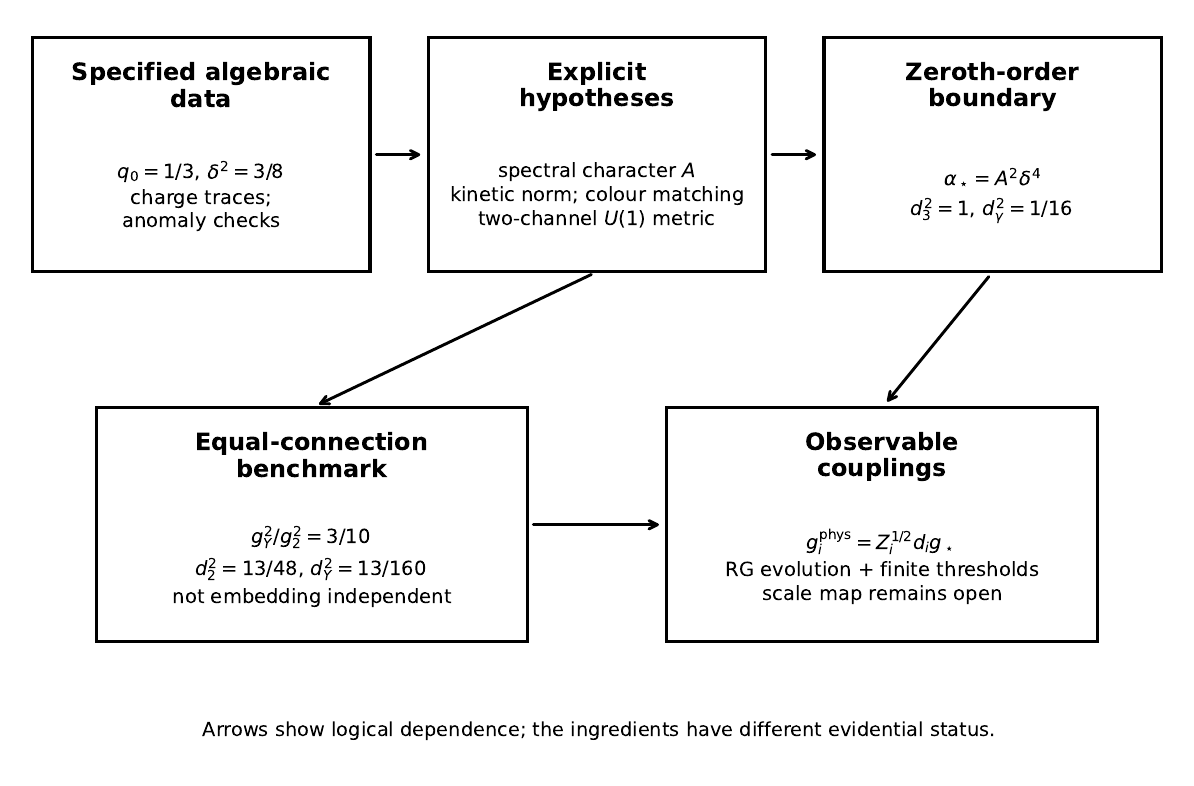}
  \caption{Logical ledger of the construction.  Exact charge traces and the adopted Jordan spectral invariant do not by themselves determine physical couplings.  The exponential character, the kinetic normalisation of the support profile, colour matching and the two-channel abelian embedding are explicit hypotheses.  Renormalisation-group evolution and finite threshold matching form a further layer.}
  \label{fig:ledger}
\end{figure}

The purpose of this revision is not to improve agreement by inserting a fitted continuous parameter.  It is to audit the normalisation steps that connect the algebraic construction to physical couplings.  In particular, the kinetic reduction and the matter-current embedding of the two abelian directions are now treated simultaneously.  The distinction between an identity within the specified construction, a result inside an effective support model, and a hypothesis about the microscopic trace dynamics will be maintained throughout.  \Cref{fig:ledger} summarises the logical architecture.

For clarity, the epistemic status of every ingredient is:
\begin{description}[style=nextline]
\item[Exact identities within the specified construction.] The charge spectrum $Q=N/3$, the one-generation trace identities, anomaly cancellation of the resulting hypercharge assignment, and the spacing $\delta^2=3/8$ of the adopted exceptional-Jordan family matrices.
\item[Derived inside stated effective models.] The direction of the invariant photon profile; the diagonal-gauge-field formula for a specified two-abelian kinetic matrix; and the weak-angle function once that matrix and the current embedding are specified.
\item[Microscopic hypotheses.] The exponential spectral character, identification of the Hilbert--Schmidt support norm with the projected kinetic norm, undiluted colour matching, the two-channel kinetic/current embedding, and identification of algebraic coefficients with boundary data near the localisation/electroweak scale.
\item[Open calculations.] Projection of the gauge Hessian and photon kinetic norm, the physical localisation spectrum and finite thresholds, and the map from Connes-time regimes to conventional momentum scale.
\end{description}

\subsection{Changes relative to the earlier versions of this paper}
\label{sec:changes}

For the reader of versions 1--3, the substantive changes are the following.
(i)~The ``democratic support vector'' hypothesis is replaced by a symmetry argument: within the linear support model the photon profile is forced by colour invariance and the real Schur lemma, so the abelian dilution $1/6$ is now a conditional theorem rather than a basis-dependent selection (\cref{sec:strong-em}).
(ii)~The colour-side matching $g_3=\gstar$ is explicitly demoted to a hypothesis.  The previously quoted cap $\alpha_s/\alpha_{\rm em}\leq8$ is now restricted to a strengthened ansatz in which each individual colour profile commutes with the complex structure; it does not follow from adjoint covariance of the gluon multiplet (\cref{sec:colour-side}).
(iii)~The electromagnetic charge trace $k_{\rm em}=8/3$ is reconstructed as one of four exact one-generation trace identities, with the state content, Weyl/Dirac counting and the role of the neutrino displayed explicitly, and with the anomaly checks that any abelian fusion must pass (\cref{sec:algebra}).
(iv)~The identification $L_P^2/L^2=3/32$ is removed from the gauge-normalisation chain, and the exponential seed is reformulated as an explicit multiplicative-character hypothesis on the charge lattice (\cref{sec:seed}).
(v)~The spinorial weak-angle derivation remains withdrawn.  The replacement two-channel model is repaired by displaying the parent generators, matter charges, link-field charges, massless and massive combinations, and canonical kinetic coefficient in one convention.  The value $3/13$ survives only as the equal-connection-coefficient benchmark, not as an embedding-independent prediction (\cref{sec:weak-completion}).
(vi)~All numerical comparisons are routed through explicit matching factors $Z_i$, the mixed-regime scale selection is stated to be phenomenological in the abstract and conclusions, and a minimal-running diagnostic quantifies the residual threshold target (\cref{sec:matching}).
(vii)~The ultraviolet running analysis is redone with the revised boundary values; the conclusion that there is no high-scale running unification is unchanged (\cref{sec:uv-running}).
(viii)~The K\'arolyh\'azy correction is retained only as a diagnostic reparametrisation: the electromagnetic datum fixes the same residual already encoded by $Z_\gamma$, and only the combination $\mu/c_K$ is identifiable (\cref{sec:karolyhazy}).
(ix)~The photon normalisation problem is reduced to explicit projected-Hessian and localisation-spectrum tests.  Existing $A_2$-profile and six-dimensional free-product calculations do not supply the required photon-specific factor (\cref{sec:open}).

\section{Trace-dynamics background and the exponential seed}
\label{sec:seed}

\subsection{The visible bosonic block}

For the trace-dynamics foundations we refer to Refs.~\cite{Adler2004,AdlerMillard1996}; for standard octonionic and exceptional-Jordan background, see Ref.~\cite{Baez2002}.

In the notation of Ref.~\cite{RajSingh2022}, the bosonic matrix variable $Q_B$ of the trace-dynamics Lagrangian has velocity
\begin{equation}
\dot Q_B=\frac{1}{L}\bigl(i\alpha_{\rm TD} q_B+L\dot q_B\bigr),
\label{eq:QBdot}
\end{equation}
where $L$ is a characteristic length and $\alpha_{\rm TD}$ is a dimensionless action parameter generated when scale invariance is broken.  To avoid a notational ambiguity in the source literature, $\alpha_{\rm TD}$ is \emph{not} the electromagnetic fine-structure constant.  The visible bosonic block is, schematically,
\begin{equation}
\mathcal L_{\mathrm{vis}}\sim \frac{L_P^2}{2L^2}\,\Tr\!\left(\dot q_B^{\dagger}+\frac{i\alpha_{\rm TD}}{L}q_B^{\dagger}\right)
\left(\dot q_B+\frac{i\alpha_{\rm TD}}{L}q_B\right).
\label{eq:Lvis}
\end{equation}
Expansion produces a pure visible block $q_B^{\dagger}q_B$, from which the photon and gluons are identified; a mixed block $q_B^{\dagger}\dot q_B$, from which the weak bosons are identified; and fermionic and boson--fermion terms not needed here.  This split is used throughout: colour and electromagnetism come from the same visible block, whereas the weak bosons are associated with the mixed sector.  The 2022 fine-structure calculation~\cite{Singh2022EPJP} identifies the dimensionless coefficient of the charged visible terms as $C\equiv\alpha_{\rm TD}^2 L_P^4/L^4$ and motivates a logarithmic dependence of $\alpha_{\rm TD}$ on the charge label.  The next subsection states the stronger functional assumption needed to turn that motivation into the exponential used here.

\subsection{The exponential seed as a character hypothesis}

The 2022 calculation introduced an exponential factor through the lower anti-down-family Jordan root~\cite{Singh2022EPJP}: the action motivates logarithmic charge dependence, while the calculation assumes $d\alpha_{\rm TD}/dq\propto\alpha_{\rm TD}$ and fixes the slope at the smallest nonzero visible charge with the corresponding Jordan eigenvalue.  A mathematically sharper formulation is to posit that the associated symmetry-breaking amplitude restricted to the additive charge lattice is a continuous multiplicative character,
\begin{equation}
 A(q_1+q_2)=A(q_1)A(q_2),\qquad A(0)=1.
 \label{eq:character}
\end{equation}
Continuity gives $A(q)=\exp(\kappa q)$.  The model-specific spectral choice is
\begin{equation}
 \kappa=\lambda_-=q_0-\delta,
 \qquad
 A:=A(q_0)=\exp\!\left[q_0(q_0-\delta)\right],
 \qquad q_0=\frac13,\quad \delta^2=\frac38,
 \label{eq:A}
\end{equation}
where $\lambda_-$ is the lower Jordan root of the $q_0$ family (\cref{sec:jordan}).  The condition $A(0)=1$ is the statement that the neutral sector returns the unbroken, scale-invariant value.  This proves the exponential form conditional on \cref{eq:character}; it does \emph{not} derive either the character law or the selection of the lower Jordan branch.  We therefore designate \cref{eq:A} as the \emph{spectral-character hypothesis}.

The common visible spectral strength used below is
\begin{equation}
 \alphastar:=A^2\delta^4
 =\frac{9}{64}\exp\!\left[\frac23\left(\frac13-\sqrt{\frac38}\right)\right]
 =0.116754180678\ldots,
 \label{eq:alphastar}
\end{equation}
This numerical value is an algebraic boundary coefficient \emph{conditional} on the character law, the lower-root selection and the use of the Jordan datum in the coupling coefficient.  None of those three inputs has yet been derived from the projected trace-dynamics action.
The corresponding parent gauge coupling is
\begin{equation}
 \gstar=\sqrt{4\pi\alphastar}=1.21127053344\ldots.
 \label{eq:gstar}
\end{equation}
The use of $\delta^4=(3/8)^2$ makes the geometric input explicitly the invariant family spacing of \cref{eq:family-spectrum} below.

\subsection{Why the length identification is removed from the gauge chain}
\label{sec:length-removal}

The original fine-structure derivation wrote the charged coefficient as $A^2(L_P^2/L^2)^2$ and set $L_P^2/L^2=3/32$ by combining half of an octonionic magnitude with an interpretation of $L$ as a Schwarzschild radius~\cite{Singh2022EPJP}.  As a universal gauge-normalisation step this is difficult to sustain: if $L=2Gm/c^2$ is particle dependent, then $L_P/L=m_P/(2m)$ is not a common order-one number for low-energy fermions.  The numerical formula itself can be retained without this identification because
\begin{equation}
 \left(\frac{3}{32}\right)^2
 =\left(\frac38\right)^2\frac1{16}
 =\delta^4\,\frac1{16}.
 \label{eq:factorization}
\end{equation}
The factor $\delta^4$ belongs to the Jordan spectral seed, whereas $1/16$ is assigned to a separate broken-phase gauge normalisation (\cref{sec:strong-em}).  This factorisation avoids conflating a particle length with a universal coupling, and it is the organising step on which the rest of the paper is built.

\subsection{Why the seed uses \texorpdfstring{$q_0=1/3$}{q0=1/3} although the electron has charge 1}
\label{sec:primitive-charge}

The seed is not attached to a particular particle species; it is attached to the minimal visible charge quantum.  In the octonionic/Furey construction the charge operator is $Q=N/3$~\cite{Furey2018}, so the visible charge lattice is $Q\in\{0,1/3,2/3,1\}$.  The electron does not carry a different primitive charge; it carries three quanta of the unit $q_0=1/3$.  Under the character law \cref{eq:character} the natural anchor is the lattice generator, and the neutral value $q=0$ returns $A=1$, consistent with the unbroken theory.  The electron is not thereby omitted: its contribution enters explicitly through the trace identities of \cref{sec:traces}, where the charged-lepton term is the final $+1$ in $\Tr Q^2$ (Dirac counting $k_{\rm em}=8/3$).  The two levels of the logic are distinct: the primitive algebraic seed is fixed by the minimal nonzero charge quantum, while the full charge content of one generation enters through the gauge normalisation traces.

\section{Algebraic data within the specified construction}
\label{sec:algebra}

\subsection{Exceptional-Jordan family spacing}
\label{sec:jordan}

For each fixed-charge family, the explicit $J_3(\OO)$ matrices used in Ref.~\cite{Singh2022EPJP} have eigenvalues
\begin{equation}
 \lambda_{q,-}=q-\delta,
 \qquad \lambda_{q,0}=q,
 \qquad \lambda_{q,+}=q+\delta,
 \qquad \delta^2=\frac38,
 \label{eq:family-spectrum}
\end{equation}
with $q\in\{0,1/3,2/3,1\}$ for the neutrino, anti-down, up and positively charged lepton families in that convention.  Equivalently, the family characteristic polynomial is
\begin{equation}
 p_q(\lambda)=(\lambda-q)\big[(\lambda-q)^2-\delta^2\big].
 \label{eq:family-poly}
\end{equation}
The quantity $\delta^2=3/8$ is therefore an invariant of the family spectrum: it is the squared half-spacing of the outer roots around the charge root.  This is a cleaner object to carry into a coupling calculation than a component norm, whose numerical value can depend on the normalisation of a chosen octonionic basis.  (In versions 1--3 the number $3/8$ was presented as the total norm of the three charged first-generation occupation sectors, $3\times1/8$; that reading is not wrong, but it inherits a basis convention, whereas the spectral spacing does not.)  We use
\begin{equation}
 q_0=\frac13,
 \qquad \lambda_-:=q_0-\delta=\frac13-\sqrt{\frac38},
 \label{eq:q0-lambda}
\end{equation}
where $q_0$ is the primitive nonzero visible charge quantum generated by the octonionic number operator $Q=N/3$~\cite{Furey2018,SinghStrongCP2026}.  No particle-specific statement is needed at this stage.

\subsection{One-generation electroweak traces}
\label{sec:traces}

We use the convention
\begin{equation}
 Q=T_L^3+Y,
 \label{eq:QTY}
\end{equation}
so that $Y$ is one half of the hypercharge denoted $Y_{\rm SM}$ in conventions where $Q=T_L^3+Y_{\rm SM}/2$.  Include one right-handed neutrino, which is neutral and does not change any trace.  Tracing over the physical chiral states of one generation gives
\begin{align}
 K_Q&:=\Tr Q^2
 =3\left[2\left(\frac23\right)^2+2\left(\frac13\right)^2\right]+2(1)^2
 =\frac{16}{3},
 \label{eq:KQ}\\
 K_2&:=\Tr (T_L^3)^2
 =3\left[\left(\frac12\right)^2+\left(-\frac12\right)^2\right]
 +\left[\left(\frac12\right)^2+\left(-\frac12\right)^2\right]
 =2,
 \label{eq:K2}\\
 K_Y&:=\Tr Y^2
 =6\left(\frac16\right)^2+3\left(\frac23\right)^2
 +3\left(\frac13\right)^2+2\left(\frac12\right)^2+1
 =\frac{10}{3},
 \label{eq:KY}\\
 K_{Y3}&:=\Tr(YT_L^3)=0.
 \label{eq:orthogonal}
\end{align}
The last identity holds doublet by doublet because $Y$ is constant across each $SU(2)_L$ doublet.  These relations also imply $K_Q=K_Y+K_2$, as required by \cref{eq:QTY,eq:orthogonal}.  The full state table is in \cref{app:traces}.

The corresponding Dirac-counting electromagnetic numerator is
\begin{equation}
 k_{\rm em}=\frac{K_Q}{2}
 =3\left[\left(\frac23\right)^2+\left(\frac13\right)^2\right]+1=\frac83,
 \label{eq:kem}
\end{equation}
the value used in the strong--electromagnetic support construction of versions 1--3, in which the factor $3$ counts colour and the final $+1$ is the charged-lepton contribution.  The trace is over the physical chiral fermion states of a single generation (equivalently, over Dirac states with the factor $2$ removed); one generation suffices because all three generations carry identical gauge charges, so a three-generation trace rescales numerator and reference non-abelian trace by the same factor and cancels from every ratio.  For the same reason Weyl versus Dirac counting cannot alter a coupling ratio when numerator and reference are counted consistently.  The neutrino is not ``omitted'': it enters all four traces and contributes zero to $K_Q$ because it is neutral.

Suppose a common parent kinetic coefficient multiplies generators normalised to equal trace.  The physical coefficients multiplying the unnormalised $Y$ and $T_L^3$ then obey
\begin{equation}
 \left.\frac{g_{Y,0}^2}{g_2^2}\right|_{\rm equal\ trace}
 =\frac{K_2}{K_Y}=\frac35,
 \qquad
 \left.\frac{e_0^2}{g^2}\right|_{\rm equal\ trace}
 =\frac{K_2}{K_Q}=\frac38.
 \label{eq:equal-trace}
\end{equation}
The second relation is the familiar tree-level unified normalisation~\cite{Slansky1981,GeorgiQuinnWeinberg1974}; it is not yet the low-energy electromagnetic coupling because the broken-phase support factor has not been included.  We emphasise what \cref{eq:equal-trace} does and does not assume: it assumes that a single parent kinetic term feeds both generators with a common trace normalisation---which is a statement about the gauge kinetic term, not merely about a matter charge sum---and this assumption is part of the effective model, to be derived eventually from the $\Tr(F_{\mu\nu}F^{\mu\nu})$ analogue of the trace-dynamics action.

\subsection{Hypercharge anomalies}
\label{sec:anomalies}

The relation $Y=Q-T_L^3$ reproduces the Standard-Model hypercharges.  Written entirely as left-handed Weyl fields, one generation is
\begin{equation}
 Q_L:(\bm3,\bm2)_{1/6},\quad
 u_R^c:(\bar{\bm3},\bm1)_{-2/3},\quad
 d_R^c:(\bar{\bm3},\bm1)_{1/3},\quad
 L_L:(\bm1,\bm2)_{-1/2},\quad
 e_R^c:(\bm1,\bm1)_1,
 \label{eq:weyl-content}
\end{equation}
plus $(\bm1,\bm1)_0$ for $\nu_R^c$.  The anomaly sums are
\begin{align}
 [SU(3)_c]^2U(1)_Y:&\quad 2\left(\frac16\right)-\frac23+\frac13=0,
 \label{eq:a33Y}\\
 [SU(2)_L]^2U(1)_Y:&\quad 3\left(\frac16\right)-\frac12=0,
 \label{eq:a22Y}\\
 [\mathrm{grav}]^2U(1)_Y:&\quad
 6\left(\frac16\right)+3\left(-\frac23\right)+3\left(\frac13\right)
 +2\left(-\frac12\right)+1=0,
 \label{eq:agravY}\\
 [U(1)_Y]^3:&\quad
 6\left(\frac16\right)^3+3\left(-\frac23\right)^3+3\left(\frac13\right)^3
 +2\left(-\frac12\right)^3+1=0.
 \label{eq:aYYY}
\end{align}
There are four $SU(2)_L$ doublets per generation after colour multiplicity, so the Witten global anomaly also cancels~\cite{Witten1982}.  These checks do not determine a coupling, but they are necessary before any proposed abelian fusion can be regarded as a viable gauging.

\section{Strong and electromagnetic sectors: the support model made precise}
\label{sec:strong-em}

\subsection{The six real ladder directions and the support space \texorpdfstring{$H_6$}{H6}}

The visible ladder operators employed in the octonionic construction are~\cite{RajSingh2022,Furey2018}
\begin{equation}
\alpha_1=\frac{-e_5+i e_4}{2},
\qquad
\alpha_2=\frac{-e_3+i e_1}{2},
\qquad
\alpha_3=\frac{-e_6+i e_2}{2}.
\label{eq:ladderops}
\end{equation}
They single out the real six-dimensional space
\begin{equation}
 H_6:=\operatorname{span}_{\RR}\{e_1,\ldots,e_6\}\;\simeq\;(\CC^3)_\RR,
 \label{eq:H6}
\end{equation}
equipped with the restriction of the standard octonionic inner product $\langle x,y\rangle=\mathrm{Re}(\bar xy)$, in which $\{e_1,\ldots,e_6\}$ is orthonormal; the complex structure $J$ defined by the ladder pairing $(e_5,e_4)$, $(e_3,e_1)$, $(e_6,e_2)$ realifies the colour fundamental $\bm 3$ of $SU(3)_c$ on $H_6$.  In the visible bosonic block, the gluon representatives are built from antisymmetric bilinears involving exactly these six directions, which is why $H_6$ is the relevant broken-phase support space for the visible gauge sector.  This addresses a definitional gap of the earlier versions: the space, the inner product and the group action are now stated objects, not operational conventions.

A remark on dimension counting, raised naturally by the adjoint: the gluons transform in the eight-dimensional adjoint of $SU(3)_c$, whereas $H_6$ carries the (realified) three-dimensional fundamental.  There is no conflict, because the support space is where the \emph{profiles} (of the surviving gauge modes and of the matter zero mode) live, not where the gauge algebra is represented; the adjoint acts on profile operators by conjugation.

\subsection{The overlap rule as a stated effective model}
\label{sec:overlap-model}

We now state the effective model explicitly, as a definition rather than as an implicit ansatz.

\begin{quote}
\textbf{Linear support model.}  Before symmetry breaking the visible sector carries a single Yang--Mills coupling $\gstar$.  After localisation, each surviving gauge mode $i$ is characterised by a self-adjoint profile operator $\widehat F_i$ on $H_6$ normalised by $\Tr\widehat F_i^2=1$, and the relevant matter zero mode by a unit vector $\eta\in H_6$.  The effective four-dimensional couplings are the linear overlaps
\begin{equation}
 g_i = g_i^{\rm parent}\,\langle\eta,\widehat F_i\,\eta\rangle,
 \label{eq:overlap-rule}
\end{equation}
where $g_i^{\rm parent}$ is the equal-trace parent coupling of the corresponding generator ($\gstar$ for colour, $e_0=\sqrt{3/8}\,\gstar$ for the electromagnetic direction, by \cref{eq:equal-trace}).
\end{quote}

This model is the analogue of a wavefunction-overlap (Kaluza--Klein-type) reduction, and it is a hypothesis: the linearity of the overlap (as opposed to an overlap squared, a trace of products of projectors, or a kinetic-term normalisation) has not been derived from the trace-dynamics Lagrangian.  Moreover, the Hilbert--Schmidt condition $\Tr\widehat F_i^2=1$ is only an auxiliary normalisation until it is shown to coincide with the norm induced by the projected four-dimensional gauge kinetic term.  Every conclusion of this section is conditional on both identifications, and \cref{sec:open} lists their derivation as an open calculation.

\subsection{The photon profile is symmetry-forced}
\label{sec:photon-forced}

In versions 1--3 the electromagnetic mode was \emph{postulated} to be the ``democratic'' vector
\[
 6^{-1/2}\sum_A\chi_A
\]
in a preferred orthonormal basis.  That formulation is open to the objection that in an abstract Euclidean six-space there is no invariant democratic vector: the object depends on the chosen basis.  The correct formulation removes the arbitrariness entirely.

The unbroken photon must not single out any colour direction: its profile must commute with the realified $SU(3)_c$ action on $H_6$.  Since the complex irreducible $\bm 3$ realifies to an irreducible real representation of complex type, the real Schur lemma gives a commutant algebra generated by the identity $\bm1_{H_6}$ and the complex structure $J$.  Self-adjointness of $\widehat F_\gamma$ removes $J$ (which is antisymmetric).  Hence, with the normalisation $\Tr\widehat F_\gamma^2=1$,
\begin{equation}
 \widehat F_\gamma=\pm\frac{\bm1_{H_6}}{\sqrt6},
 \label{eq:Fgamma}
\end{equation}
and the magnitude of the overlap with \emph{every} unit matter profile is $6^{-1/2}$, independently of any basis and of the location of $\eta$.  Combining this support dilution with the equal-trace electromagnetic factor in \cref{eq:equal-trace} gives
\begin{equation}
 \frac{e^2}{\gstar^2}=\frac38\times\frac16=\frac1{16}.
 \label{eq:em16}
\end{equation}
The real Schur argument is a theorem \emph{inside the stated representation-theoretic assumptions}: it fixes the invariant profile direction, not its physical coupling normalisation.  Equation~\eqref{eq:em16} additionally assumes the linear overlap rule, equal-trace parent normalisation, and equality of the Hilbert--Schmidt and projected kinetic norms.  A rescaling of the profile can otherwise be compensated by a rescaling of the parent coupling.  The basis ambiguity has therefore been removed, but the kinetic-normalisation problem has not.

\subsection{The colour side: adjoint covariance and \texorpdfstring{$d_3=1$}{d3=1} as a hypothesis}
\label{sec:colour-side}

The colour side is less secure, and we state exactly why.  Saturating $g_3=\gstar$ in \cref{eq:overlap-rule} requires $\langle\eta,\widehat F_c\,\eta\rangle=1$, for example the rank-one profile $\widehat F_c=|\eta\rangle\langle\eta|$.  If one imposes the \emph{additional} condition $[\widehat F_c,J]=0$ on each individual colour profile, then a self-adjoint profile has even rank.  The minimal such choice is $\widehat F_c=P_2/\sqrt2$, with $P_2$ the projector onto the complex line (real two-plane) containing $\eta$.  Its overlap is $1/\sqrt2$, giving $g_3^2=\gstar^2/2$ and the conditional bound
\begin{equation}
 \left.\frac{\alpha_s}{\alpha_{\rm em}}\right|_{\text{invariant linear model}}
 \le\;\frac{1/2}{1/16}=8 .
 \label{eq:cap8}
\end{equation}
This is not a model-independent obstruction.  A photon is a colour singlet and its profile should commute with the colour action.  An individual gluon component instead belongs to an adjoint-covariant eight-dimensional family; covariance of the family under conjugation does not imply $[\widehat F_c,J]=0$ component by component.  Hence \cref{eq:cap8} applies only to the strengthened class of individually $J$-invariant colour-profile ans\"atze.  The projected localisation action has not yet selected either that class or an alternative covariant family.  We accordingly retain
\begin{equation}
 d_3=1,
 \qquad g_3^{(0)}=\gstar,
 \label{eq:colour-match}
\end{equation}
only as a \emph{strong-sector matching hypothesis}.  A microscopic localisation mechanism must derive the covariant eight-profile family and explain whether the colour mode is diluted while preserving exact $SU(3)_c$; this is listed as an open problem in \cref{sec:open}.

With that caveat,
\begin{align}
 \alpha_3^{(0)}&=\alphastar=0.116754180678\ldots,
 \label{eq:a3zero}\\
 \alpha_\gamma^{(0)}&=\frac{\alphastar}{16}
 =0.007297136292\ldots,
 \qquad [\alpha_\gamma^{(0)}]^{-1}=137.04006064\ldots.
 \label{eq:agzero}
\end{align}
The superscript $(0)$ is deliberate: these are algebraic boundary coefficients before conventional running and finite threshold matching.  As a conditional statement, the full chain reads:
\begin{quote}\itshape
If the broken phase is described by the linear support model of \cref{sec:overlap-model} with the equal-trace parent normalisation \cref{eq:equal-trace}, then colour invariance forces the support contribution to $e^2/\gstar^2$ to be $1/6$ and hence $e^2/\gstar^2=1/16$; if in addition the colour mode matches undiluted ($d_3=1$), then $\alpha_s/\alpha_{\rm em}=16$ at the algebraic boundary.
\end{quote}
This is the precise form of the claim that earlier versions stated more loosely.

\section{A normalisation no-go for the earlier weak-angle equation}
\label{sec:weak-nogo}

The 2022 weak-angle construction made three linked moves~\cite{RajSingh2022}: a half-angle rotation on octonionic spinor space was applied to the bosonic pair $(B,W^3)$; in an asymptotic limit the fields were identified with $Y^2$ and $(T_L^3)^2$; square roots were then inserted into $Q=Y+T_L^3$.  The resulting equation, $1=\tfrac12\sqrt{\cos(\theta_W/2)}+\sqrt{\sin(\theta_W/2)}$, gave $\sin^2\theta_W\simeq0.2497$.

There are four independent problems.

\paragraph{Canonical rescaling.}
Before kinetic terms are canonically normalised one may rescale
\begin{equation}
 B_\mu\mapsto c_B B_\mu,
 \qquad W^3_\mu\mapsto c_W W^3_\mu,
 \label{eq:rescale}
\end{equation}
with compensating changes in $g_Y$ and $g_2$.  A physical mixing angle depends on the canonically normalised couplings, $\tan\theta_W=g_Y/g_2$, and is invariant under such bookkeeping.  The square-root field equation changes when $c_B/c_W$ changes; indeed the source paper explicitly noted that the limiting constants multiplying $B$ and $W^3$ need not be equal.  The numerical root is therefore not normalisation invariant.

\paragraph{Linearity of gauge couplings.}
Gauge connections couple linearly,
\begin{equation}
 D_\mu=\partial_\mu-i g_2 T_L^a W^a_\mu-i g_Y YB_\mu.
 \label{eq:covder}
\end{equation}
Identifying a connection with a squared generator and then taking a square root is not a gauge-covariant operation.  It discards the sign of $Y$ and $T_L^3$ and cannot be representation independent.

\paragraph{Spinor versus adjoint angle.}
A spinor transforms with a half-angle because $\mathrm{Spin}(n)$ double-covers $SO(n)$.  A gauge connection transforms in an adjoint or affine representation.  A half-angle can enter a bosonic field relation only after an explicit intertwining map from the spinor construction to the canonically normalised gauge kinetic terms has been proved.  Such a map was not supplied.

\paragraph{Scheme.}
When $g_Y$ and $g_2$ are defined in the $\MSbar$ scheme, their ratio determines the running $\MSbar$ angle.  This is distinct from the on-shell definition $1-M_W^2/M_Z^2$ and from the effective leptonic angle; conversion requires radiative corrections~\cite{Sirlin1980,PDG2024}.  A gauge-coupling derivation must state the scheme and scale before comparison.

For these reasons the earlier equation is retained only as historical motivation.  It is not used in any subsequent formula in this paper, and the value $0.2497$ is no longer presented as a prediction of the programme.

\section{Current-consistent two-\texorpdfstring{$U(1)$}{U(1)} reduction}
\label{sec:weak-completion}

\subsection{Why an additional abelian matching step is required}

The current $\EomE$ construction recovers the physical hypercharge assignment as the consistency relation
\begin{equation}
 Y=Q-T_L^3,
 \qquad Q=N/3,
 \label{eq:Yderived}
\end{equation}
not as a representation-independent rescaling of the $T_L^8$ Cartan of an electroweak $SU(3)$~\cite{SinghStrongCP2026,SinghBook2026}.  The Stage-1 Cartan has the correct Higgs-doublet quantum number but annihilates right-handed fermions in the two-sided assignment, whereas physical hypercharge does not.  Consequently the framework owes a leaf-level fusion of a Cartan abelian direction with the Cl$(6)$ trace/number direction, together with a fate for the orthogonal combination.

This structural debt suggests a controlled place at which a coupling reduction can occur.  It does not license an arbitrary numerical factor: the parent generators, matter current, link-field charges and kinetic matrix must be specified together.

\subsection{Two-channel diagonal locking}
\label{sec:locking}

Consider two abelian parent connections $A^{(1)}$ and $A^{(2)}$.  Allow the gauge-invariant abelian kinetic mixing of Ref.~\cite{Holdom1986}.  In a convention where the gauge couplings occur in the kinetic terms, take
\begin{equation}
 \mathcal L_{\rm kin}
 =-\frac14\left[
 \frac{F_1^2}{g_1^2}+\frac{F_2^2}{g_2'^2}
 +\frac{2\chi}{g_1g_2'}F_1^{\mu\nu}F_{2\mu\nu}
 \right].
 \label{eq:twoU1kin}
\end{equation}
For every Standard-Model multiplet $f$, write the two parent generators as
\begin{equation}
 q_1(f)=c_1Y_f,\qquad q_2(f)=c_2Y_f,\qquad c_1+c_2=1.
 \label{eq:parentcharges}
\end{equation}
For completeness, \cref{tab:parentcharges} displays the action on every chiral
Standard-Model matter multiplet.  The same constants $c_1,c_2$ apply to all
multiplets; they are not representation-dependent fitting parameters.
\begin{table}[H]
\centering
\caption{Parent abelian charges in the convention $Q=T_L^3+Y$.  Colour and weak multiplicities are suppressed.}
\label{tab:parentcharges}
\begin{tabular}{@{}lccc@{}}
\toprule
multiplet $f$ & $Y_f$ & $q_1(f)=c_1Y_f$ & $q_2(f)=c_2Y_f$ \\
\midrule
$q_L$ & $1/6$ & $c_1/6$ & $c_2/6$ \\
$u_R$ & $2/3$ & $2c_1/3$ & $2c_2/3$ \\
$d_R$ & $-1/3$ & $-c_1/3$ & $-c_2/3$ \\
$\ell_L$ & $-1/2$ & $-c_1/2$ & $-c_2/2$ \\
$e_R$ & $-1$ & $-c_1$ & $-c_2$ \\
\bottomrule
\end{tabular}
\end{table}
The complete pre-locking matter coupling is
\begin{equation}
 \mathcal L_{\rm int}=J_Y^\mu\bigl(c_1A_\mu^{(1)}+c_2A_\mu^{(2)}\bigr).
 \label{eq:parentcurrent}
\end{equation}
A link scalar $\Phi$ of charges $(+q_\ell,-q_\ell)$ has
\begin{equation}
 D_\mu\Phi=\bigl[\partial_\mu-iq_\ell(A_\mu^{(1)}-A_\mu^{(2)})\bigr]\Phi.
 \label{eq:linkcov}
\end{equation}
As in diagonal breaking of a product gauge group~\cite{ArkaniHamed2001}, its nonzero vacuum expectation value gives a mass to the difference connection $A^{(1)}-A^{(2)}$ and leaves the diagonal branch
\begin{equation}
 A^{(1)}=A^{(2)}=A_Y.
 \label{eq:diagbranch}
\end{equation}
Equation~\eqref{eq:parentcurrent} then becomes $\mathcal L_{\rm int}=J_Y^\mu A_{Y\mu}$: the surviving generator is exactly the physical Standard-Model $Y$, rather than $2Y$.  Each parent and every mixed abelian anomaly is proportional to the corresponding anomaly sum for $Y$ and therefore vanishes; the scalar link introduces no chiral anomaly.  The orthogonal connection is massive, while its detailed canonical direction depends on the kinetic matrix.

Substitution of \cref{eq:diagbranch} into \cref{eq:twoU1kin} gives
\begin{equation}
 \frac1{g_Y^2}
 =\frac1{g_1^2}+\frac1{g_2'^2}+\frac{2\chi}{g_1g_2'}.
 \label{eq:harmonic}
\end{equation}
After defining the canonical massless field $B_Y=A_Y/g_Y$, the current is $g_YJ_Y^\mu B_{Y\mu}$, so \cref{eq:harmonic} is simultaneously the kinetic and current normalisation.  Let $g_{Y,0}$ be the coefficient inferred for the full generator $Y$ from \cref{eq:equal-trace}, and define
\begin{equation}
 g_Y^2=\eta\,g_{Y,0}^2.
 \label{eq:eta-def}
\end{equation}
There are now two inequivalent normalisation hypotheses.

\paragraph{Fixed connection coefficients.}
If the microscopic action assigns the same coefficient to the two parent \emph{connections}, independently of how the physical current is partitioned,
\begin{equation}
 g_1=g_2'=g_{Y,0},\qquad \chi=0
 \quad\Longrightarrow\quad
 \eta=\frac12.
 \label{eq:eta-half}
\end{equation}
This is an internally consistent benchmark because \cref{eq:parentcharges} has already repaired the current.  It is nevertheless an additional microscopic hypothesis: equality of the connection coefficients does not follow from applying the charge trace separately to the scaled generators $c_iY$.

\paragraph{Trace-induced coefficients.}
If instead the same trace prescription that gives \cref{eq:equal-trace} is applied separately to $q_i=c_iY$, then, at zero kinetic mixing,
\begin{equation}
 \frac1{g_i^2}=\frac{c_i^2}{g_{Y,0}^2},
 \qquad
 \eta=\frac1{c_1^2+c_2^2}.
 \label{eq:traceinduced}
\end{equation}
For nonnegative $c_i$ with $c_1+c_2=1$, this gives $1\leq\eta\leq2$; equal splitting gives $\eta=2$, not $1/2$.  The endpoint $\eta=1$ is the limiting single-channel case: a strictly positive-definite two-channel trace-induced kinetic matrix requires $c_1c_2\ne0$, for which $1<\eta\leq2$ at zero mixing.  With a dimensionless cross coefficient $\rho$ in the trace-induced kinetic matrix, the denominator becomes $c_1^2+c_2^2+2\rho c_1c_2$.  For $c_1c_2\ne0$, positivity implies $|\rho|<1$, and still excludes $\eta=1/2$ for a nonnegative partition.  Thus the factor $1/2$ is not invariant under changing between fixed-connection and trace-induced normalisations.  The projected microscopic action must decide which kinetic metric is correct.

A possible objection must be addressed.  The gravi-weak and internal $SU(2)_L$ labels are also fused in the broader construction.  Why does the same factor not automatically alter $g_2^2$?  In the present benchmark the $SU(2)_L$ fusion is an identification on one surviving index space, with only one propagating leaf-level kinetic term retained; the two abelian directions instead remain independent kinetic directions before the orthogonal combination is removed.  This distinction is a hypothesis about the localisation map.  If the microscopic action contains two independent surviving $SU(2)_L$ kinetic terms with the same diagonal reduction as hypercharge, any common factor cancels from the ratio.  This is a sharp test for the full action.

\subsection{Weak mixing angle}

Combining \cref{eq:equal-trace,eq:eta-def} gives, for any fully specified kinetic/current embedding,
\begin{equation}
 r_W(\eta):=\frac{g_Y^2}{g_2^2}=\frac{3\eta}{5}.
 \label{eq:rWeta}
\end{equation}
The canonically defined weak angle is then
\begin{equation}
 \sin^2\theta_{W,\star}(\eta)
 =\frac{g_Y^2}{g_2^2+g_Y^2}
 =\frac{3\eta}{5+3\eta}.
 \label{eq:s2eta}
\end{equation}
For the trace-induced prescription at zero kinetic mixing, $c_i\geq0$ and $c_1+c_2=1$ instead imply
\[
 \frac38\leq\sin^2\theta_{W,\star}=\frac{3\eta}{5+3\eta}\leq\frac6{11},
\]
with the lower endpoint understood as the single-channel limit and equal splitting giving $6/11$.  This range is far from the quoted electroweak value and does not provide another near-agreement.

For the fixed-connection benchmark \cref{eq:eta-half},
\begin{equation}
 \boxed{
 \frac{g_Y^2}{g_2^2}=\frac3{10},
 \qquad
 \sin^2\theta_{W,\star}=\frac3{13}=0.230769230769\ldots}
 \label{eq:3over13}
\end{equation}
No square root of a gauge field and no half-angle assignment to a bosonic connection is involved.  However, \cref{eq:traceinduced} proves that this number is not an embedding-independent consequence of equal-trace normalisation.  We therefore call $3/13$ an \emph{equal-connection benchmark}, not a prediction, even at zeroth order.

\begin{figure}[t]
  \centering
  \includegraphics[width=0.82\textwidth]{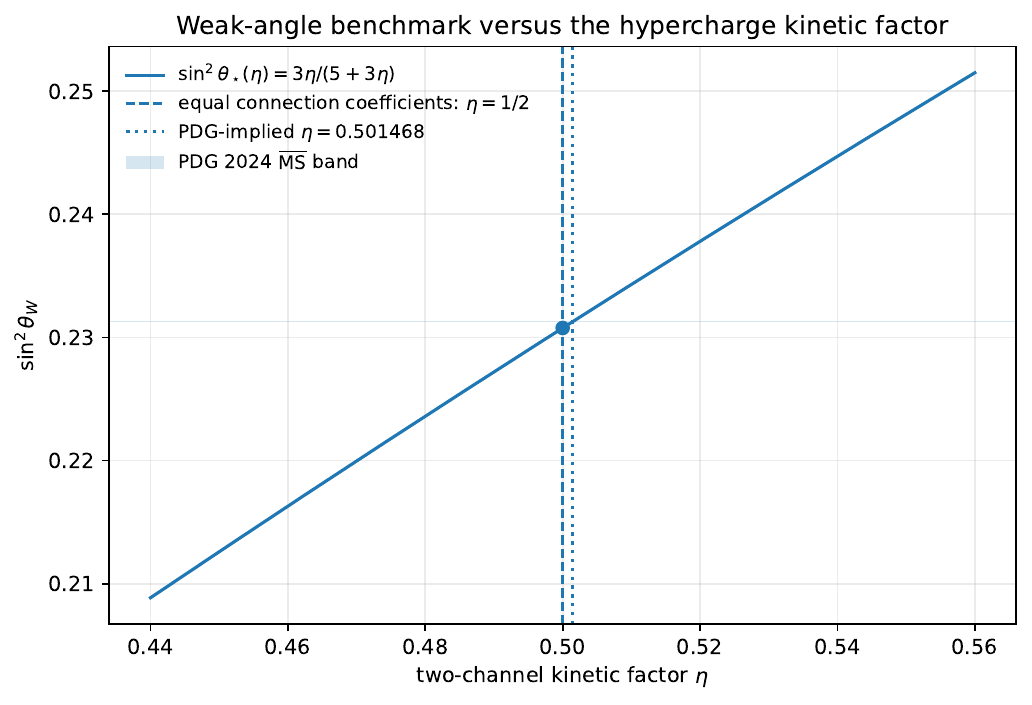}
  \caption{Weak-angle boundary as a function of the two-channel factor $\eta$.  The equal-connection benchmark gives $\eta=1/2$ and $\sin^2\theta_{W,\star}=3/13$.  The narrow band is the PDG 2024 $\MSbar$ uncertainty at $M_Z$.  If the benchmark is treated as exact, the difference is about $13\sigma$; no uncomputed theory effect is assigned as a statistical error.}
  \label{fig:eta}
\end{figure}

The PDG 2024 electroweak reference value used here is $\sin^2\widehat\theta_W(M_Z)=0.23129\pm0.00004$ in the $\MSbar$ scheme~\cite{Sirlin1980,PDG2024}.  Solving \cref{eq:s2eta} for the value selected by this datum gives
\begin{equation}
 \eta_{\rm exp}
 =\frac{5\sin^2\widehat\theta_W}{3(1-\sin^2\widehat\theta_W)}
 =0.501468\pm0.000113,
 \label{eq:etaexp}
\end{equation}
which is $(0.294\pm0.023)\%$ above $1/2$.  Within the fixed-connection benchmark only, this can be represented by a small kinetic mixing
\begin{equation}
 \eta=\frac1{2(1+\chi)}
 \quad\Longrightarrow\quad
 \chi_{\rm exp}=-0.002927\pm0.000224.
 \label{eq:chiexp}
\end{equation}
Equations~\eqref{eq:etaexp} and \eqref{eq:chiexp} are data-inferred targets, not fits to be inserted into the construction.  If the equal-connection benchmark is treated as exact with no theory error, $3/13$ lies about $13$ experimental standard deviations below the quoted central value.  The incompleteness of the microscopic model is not a statistical uncertainty and is not used to reduce this pull.  \Cref{fig:eta} shows the sensitivity.

\section{A conditional equal-connection coupling ledger}
\label{sec:ledger}

Write every algebraic boundary coupling as
\begin{equation}
 g_i^{(0)}=d_i\gstar,
 \qquad \alpha_i^{(0)}=d_i^2\alphastar.
 \label{eq:di}
\end{equation}
The strong and electromagnetic sectors give
\begin{equation}
 d_3^2=1,
 \qquad d_\gamma^2=\frac1{16}.
 \label{eq:d3dgamma}
\end{equation}
The electroweak relation
\begin{equation}
 \frac1{e^2}=\frac1{g_2^2}+\frac1{g_Y^2}
 \label{eq:ewharmonic}
\end{equation}
combined with the \emph{benchmark} $g_Y^2/g_2^2=3/10$ fixes $d_2$ and $d_Y$ rather than leaving an independent weak normalisation:
\begin{equation}
 \frac{d_Y^2}{d_2^2}=\frac3{10},
 \qquad
 \frac{d_2^2d_Y^2}{d_2^2+d_Y^2}=\frac1{16}.
 \label{eq:dsolve}
\end{equation}
The unique positive solution within that benchmark is
\begin{equation}
 \boxed{
 d_3^2=1,
 \qquad d_2^2=\frac{13}{48},
 \qquad d_Y^2=\frac{13}{160},
 \qquad d_\gamma^2=\frac1{16}.}
 \label{eq:dledger}
\end{equation}
Equivalently,
\begin{equation}
 \alpha_3^{(0)}:\alpha_2^{(0)}:\alpha_Y^{(0)}:\alpha_\gamma^{(0)}
 =16:\frac{13}{3}:\frac{13}{10}:1.
 \label{eq:ratioledger}
\end{equation}
The resulting numbers are collected in \cref{tab:zeroth}.

\begin{table}[t]
\centering
\caption{Conditional equal-connection boundary ledger from \cref{eq:alphastar,eq:dledger}.  The weak entries depend on the kinetic hypothesis \cref{eq:eta-half}; the entries are not all asserted at the same conventional momentum scale.}
\label{tab:zeroth}
\begin{tabular}{@{}lccc@{}}
\toprule
sector $i$ & $d_i^2$ & $\alpha_i^{(0)}$ & $g_i^{(0)}$ \\
\midrule
$SU(3)_c$ & $1$ & $0.116754180678$ & $1.21127053344$ \\
$SU(2)_L$ & $13/48$ & $0.031620923934$ & $0.63036517140$ \\
$U(1)_Y$ & $13/160$ & $0.009486277180$ & $0.34526522384$ \\
$U(1)_{\rm em}$ & $1/16$ & $0.007297136292$ & $0.30281763336$ \\
\bottomrule
\end{tabular}
\end{table}

A useful conceptual point follows.  The weak coupling is not set equal to the colour coupling.  The visible $q_B^\dagger q_B$ block and the mixed $q_B^\dagger\dot q_B$ block have different support and localisation data in the trace-dynamics Lagrangian~\cite{RajSingh2022,SinghBook2026}; \cref{eq:dledger} parametrises that distinction under the equal-connection hypothesis instead of deriving it from the present microscopic action.

\section{From algebraic boundary data to observables}
\label{sec:matching}

\subsection{Matching factors}

To avoid mixing scales silently, define
\begin{equation}
 g_i^{\rm phys}(\mu)=Z_i^{1/2}(\mu,\mu_\star)\,d_i\gstar,
 \label{eq:Zdef}
\end{equation}
where $Z_i$ includes ordinary running from a localisation/matching regime $\mu_\star$, finite threshold corrections, and any conversion between the octonionic regime label and conventional momentum subtraction.  This equation is a ledger, not a claim that all effects factorise microscopically.

Using the 2022 CODATA value $\alpha(0)=0.0072973525643(11)$~\cite{CODATA2022} and the PDG world average $\alpha_s(M_Z)=0.1180\pm0.0009$~\cite{PDG2024}, the strong and electromagnetic squared-coupling factors are
\begin{align}
 Z_\gamma(0)&=\frac{\alpha(0)}{\alpha_\gamma^{(0)}}
 =1.0000296379,
 \label{eq:Zgamma}\\
 Z_3(M_Z)&=\frac{\alpha_s(M_Z)}{\alpha_3^{(0)}}
 =1.0106704472.
 \label{eq:Z3}
\end{align}
Thus the visually striking agreement of the fine-structure expression corresponds to a $2.96\times10^{-5}$ fractional correction, but this is vastly larger than the metrological uncertainty ($1.5\times10^{-10}$) and must not be described as precision agreement.  At present $Z_\gamma$ is defined by the measured value in \cref{eq:Zgamma}; it has not been calculated from a photon localisation spectrum.  Equivalently, the required kinetic-coefficient shift is $K_\gamma^{\rm phys}/K_\gamma^{(0)}=Z_\gamma^{-1}=0.99997036296\ldots$.  The strong coefficient needs a $1.07\%$ correction in $g_3^2$.

At $M_Z$, use $\widehat\alpha^{(5)}(M_Z^2)^{-1}=127.930\pm0.008$ and $\sin^2\widehat\theta_W=0.23129$~\cite{PDG2024}.  Then
\begin{equation}
 \alpha_2^{\rm exp}(M_Z)=\frac{\widehat\alpha(M_Z)}{\sin^2\widehat\theta_W},
 \qquad
 \alpha_Y^{\rm exp}(M_Z)=\frac{\widehat\alpha(M_Z)}{1-\sin^2\widehat\theta_W},
 \label{eq:alpha2Yexp}
\end{equation}
so that
\begin{equation}
 Z_2(M_Z)=1.06879937,
 \qquad
 Z_Y(M_Z)=1.07193699,
 \qquad
 \frac{Z_Y}{Z_2}=1.00293565.
 \label{eq:Zweak}
\end{equation}
Within the equal-connection benchmark, most of the common $\simeq7\%$ increase from the boundary electromagnetic coefficient to the $M_Z$ electroweak couplings is compatible with ordinary running.  The relative factor $Z_Y/Z_2$ is the data-inferred $0.294\%$ shift quantified by \cref{eq:etaexp}.  The tree-level identity \cref{eq:ewharmonic} may be differentiated algebraically for inverse couplings in an unbroken-phase scheme, but this does not mean that the physical low-energy QED coupling runs with coefficient $b_Y+b_2$ across the electroweak threshold.  Matching from $SU(2)_L\times U(1)_Y$ to QED requires electroweak threshold corrections and a declared scheme~\cite{Sirlin1980,PDG2024}.

\subsection{Same-scale versus mixed-regime ratios}
\label{sec:mixed-regime}

The algebraic ratio is $\alpha_3^{(0)}/\alpha_\gamma^{(0)}=16$.  Two different experimental comparisons must not be conflated:
\begin{align}
 \frac{\alpha_s(M_Z)}{\widehat\alpha^{(5)}(M_Z^2)}&=15.096\pm0.115,
 \label{eq:samescale}\\
 \frac{\alpha_s(M_Z)}{\alpha(0)}&=16.170\pm0.123.
 \label{eq:mixedscale}
\end{align}
The first is the conservative same-scale benchmark and differs from $16$ by about $6\%$ ($7.9$ experimental standard deviations if theory uncertainty is set to zero).  The second is a mixed-regime comparison and differs by $1.05\%$ ($1.4\sigma$).  The octonionic interpretation motivates different asymptotic readings of the abelian and strongly self-interacting sectors---the flat low-energy $U(1)$ mode versus the broken-phase colour block---but it has not yet derived why those readings should map specifically to $Q=0$ and $Q=M_Z$.  \emph{The selection of these two regimes is therefore a phenomenological choice motivated by the framework, not a derived renormalisation-group statement}; this is stated here, in the abstract, and in the conclusions.  The matching factors in \cref{eq:Zdef} are the appropriate place to keep this unresolved issue visible, and the same-scale benchmark \cref{eq:samescale} is retained throughout as the conservative comparison.

\subsection{Minimal-running diagnostic near the electroweak scale}
\label{sec:running}

As a diagnostic, take the measured $\MSbar$ electroweak couplings at $M_Z$ and evolve them at one loop with the one-Higgs-doublet Standard Model,
\begin{equation}
 \frac{1}{g_i^2(\mu)}=\frac{1}{g_i^2(M_Z)}-\frac{b_i}{8\pi^2}\ln\frac{\mu}{M_Z},
 \qquad b_Y=\frac{41}{6},\quad b_2=-\frac{19}{6}.
 \label{eq:EWrun}
\end{equation}
Using those measured couplings as inputs, the diagnostic scale at which $g_Y^2/g_2^2=3/10$ is
\begin{equation}
 \mu_{\rm weak}=82.155\ \mathrm{GeV}.
 \label{eq:muweak}
\end{equation}
With two Higgs doublets the value changes only to $81.959$ GeV at this order.  A two-loop gauge-running cross-check using the Machacek--Vaughn gauge coefficients~\cite{MachacekVaughn1983}, with $y_t(M_Z)=0.99$ held fixed over this short interval, shifts the one-doublet value to $82.024$ GeV, a $0.16\%$ change in scale.  This is not a full coupled two-loop Standard-Model evolution: the Yukawa and scalar couplings are not evolved and no electroweak threshold matching is included.

\begin{figure}[t]
  \centering
  \includegraphics[width=0.86\textwidth]{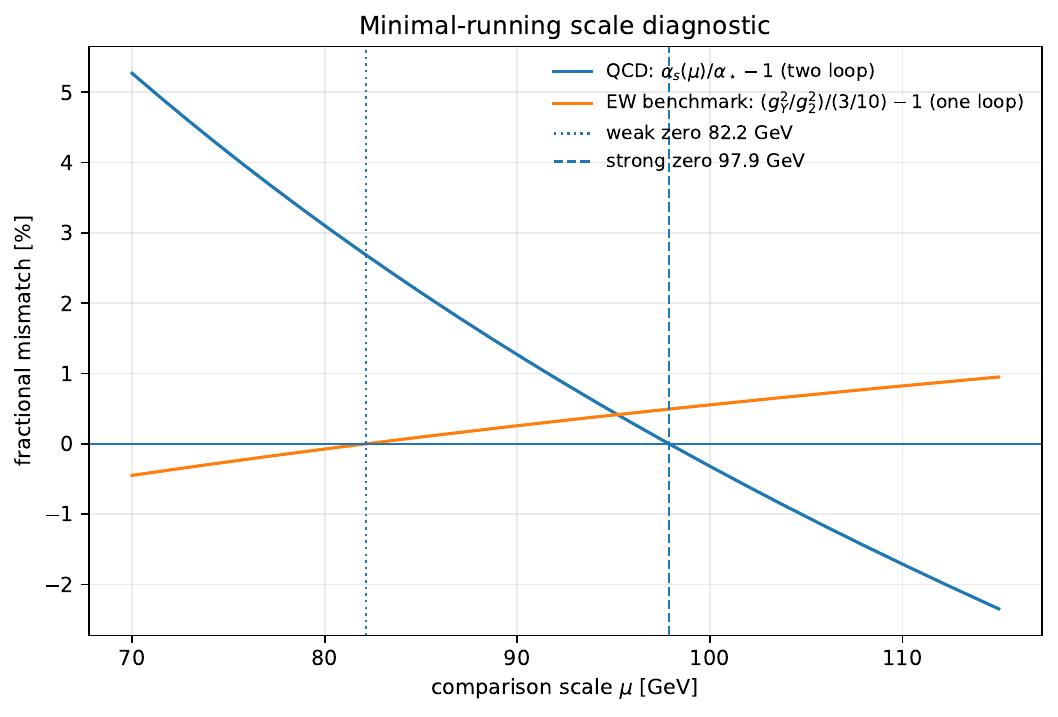}
  \caption{Scale diagnostic using measured couplings as inputs.  Minimal running places the weak benchmark $3/10$ at $82.2$ GeV and the strong coefficient $\alphastar$ at $97.9$ GeV.  The two-loop gauge-running and three-loop QCD cross-checks shift these roots only slightly.  They are data-driven roots, not octonionic predictions.}
  \label{fig:scale}
\end{figure}

For QCD, evolve at two loops with $n_f=5$,
\begin{equation}
 \frac{d\alpha_s}{d\ln\mu}
 =-\frac{\beta_0}{2\pi}\alpha_s^2
 -\frac{\beta_1}{8\pi^2}\alpha_s^3,
 \quad
 \beta_0=11-\frac{2n_f}{3},\quad
 \beta_1=102-\frac{38n_f}{3}.
 \label{eq:QCDrun}
\end{equation}
The scale at which the measured coupling equals $\alphastar$ is
\begin{equation}
 \mu_{\rm strong}=97.876\ \mathrm{GeV},
 \label{eq:mustrong}
\end{equation}
moving only to $97.862$ GeV at three loops~\cite{LarinVermaseren1993}.  The two conditions therefore do not identify an exact common scale under minimal running.  A least-squares compromise near $97.29$ GeV leaves a $0.090\%$ strong mismatch and a $0.476\%$ weak-ratio mismatch.  \Cref{fig:scale} visualises the result.

The residual can be stated as a concrete threshold target.  At $\mu_{\rm strong}$ the measured ratio overshoots, $(g_Y^2/g_2^2)/(3/10)-1=+0.493\%$; restoring the algebraic condition there requires a single finite matching contribution
\begin{equation}
 \Delta\!\left(\frac1{g_Y^2}\right)=+0.0386,
 \qquad\text{i.e.}\qquad
 \frac{\Delta(1/g_Y^2)}{1/g_Y^2}=+0.49\%,
 \label{eq:thresholdtarget}
\end{equation}
or the equivalent split between the two inverse squared couplings.  In one-loop language $\Delta(1/g_Y^2)=(\Delta b_Y/8\pi^2)\ln(M/\mu_{\rm strong})$, so \cref{eq:thresholdtarget} corresponds to $\Delta b_Y\ln(M/\mu_{\rm strong})\simeq3.0$.  This is compatible in order of magnitude with an ordinary one-loop threshold from a modest number of suitably charged states, but its sign and magnitude are spectrum dependent.  The correction is not zero and should not be concealed.

\subsection{Ultraviolet running: no high-scale unification}
\label{sec:uv-running}

A conventional grand-unification reading of the $\EomE$ structure would require the three couplings to meet under renormalisation-group flow.  We repeat the ultraviolet diagnostic of version 3 with the \emph{revised} boundary values, since the earlier exercise used the now-withdrawn weak angle.  Assign the ledger outputs to a boundary set at $M_Z$---with the explicit caveat that the electromagnetic entry is really the asymptotic $Q\to0$ value---
\begin{equation}
 \alpha_{\rm em}=\alpha_\gamma^{(0)}=0.00729713629,\qquad
 \sin^2\theta_{W,\star}=\frac3{13},\qquad
 \alpha_s=\alphastar,
 \label{eq:uv-inputs}
\end{equation}
which in canonical $SU(5)$ normalisation, $\alpha_1\equiv(5/3)\alpha_Y$~\cite{Slansky1981,GeorgiQuinnWeinberg1974}, gives
\begin{equation}
 \alpha_1^{-1}(M_Z)=63.25,\qquad
 \alpha_2^{-1}(M_Z)=31.62,\qquad
 \alpha_3^{-1}(M_Z)=8.57 .
 \label{eq:uv-bc}
\end{equation}
Note that in this normalisation the equal-trace relation \cref{eq:equal-trace} would give $\alpha_1=\alpha_2$ at the boundary; the two-channel factor $\eta=1/2$ is what separates them, and the framework-normalised statement ``$(10/3)\alpha_Y=\alpha_2$'' is precisely the condition whose measured-coupling zero is $\mu_{\rm weak}=82.2$ GeV in \cref{eq:muweak}.

\begin{figure}[t]
  \centering
  \includegraphics[width=0.8\textwidth]{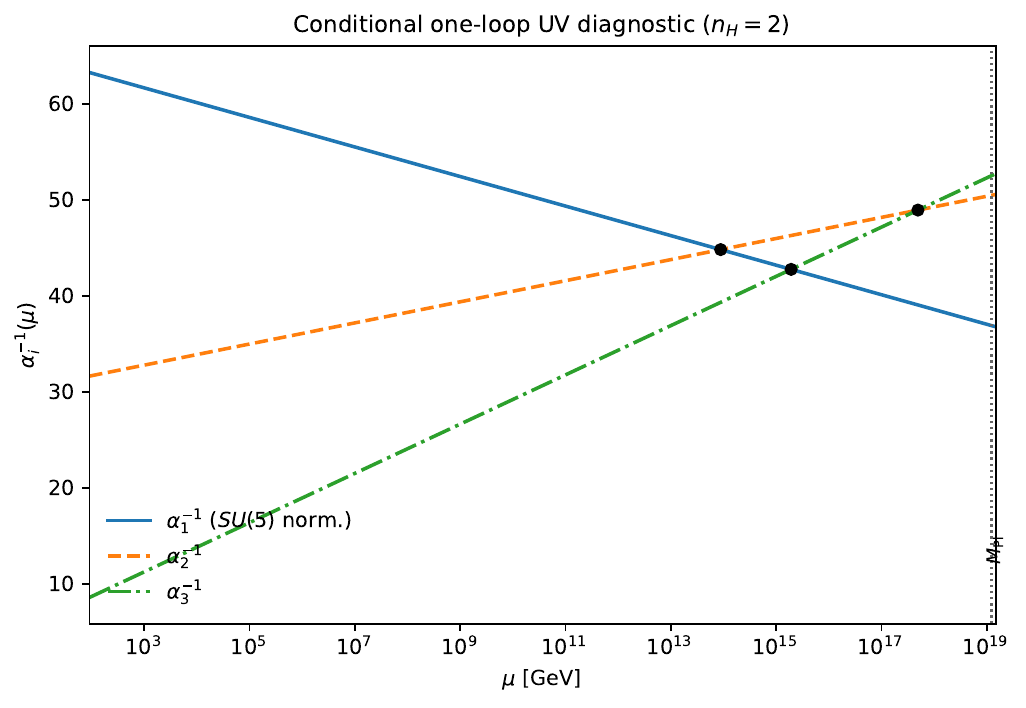}
  \caption{One-loop ultraviolet evolution of the conditional equal-connection benchmark with the minimal desert spectrum ($n_H=2$), canonical $SU(5)$ normalisation.  The electromagnetic entry is the asymptotic $Q\to0$ value, placed at $M_Z$ only for this running diagnostic; it is not a physical same-scale matching condition.  The three pairwise crossings of \cref{eq:uv-crossings} are marked; there is no common meeting point, and the conclusion is unchanged with $n_H=1$ or with fully measured boundary inputs.}
  \label{fig:uv}
\end{figure}

For the desert spectrum we take the Standard-Model gauge-charged fields and one additional electroweak scalar doublet, giving $n_H=2$ above $M_Z$, together with sterile neutrinos as gauge singlets.  Thus $(b_1,b_2,b_3)=(21/5,-3,-7)$.  One-loop evolution of \cref{eq:uv-bc} gives pairwise crossings at
\begin{equation}
\begin{aligned}
 \alpha_1^{-1}=\alpha_2^{-1}:&\quad \mu\simeq8.8\times10^{13}\ \mathrm{GeV},
 \qquad \alpha^{-1}\simeq44.8,\\
 \alpha_1^{-1}=\alpha_3^{-1}:&\quad \mu\simeq1.9\times10^{15}\ \mathrm{GeV},
 \qquad \alpha^{-1}\simeq42.7,\\
 \alpha_2^{-1}=\alpha_3^{-1}:&\quad \mu\simeq4.9\times10^{17}\ \mathrm{GeV},
 \qquad \alpha^{-1}\simeq48.9,
\end{aligned}
\label{eq:uv-crossings}
\end{equation}
spread over roughly four orders of magnitude in scale; at $M_{\rm Pl}=1.22\times10^{19}$ GeV the inverse couplings are $(36.9,\,50.5,\,52.5)$, a spread $\Delta\alpha^{-1}\simeq15.6$.  See \cref{fig:uv}.  With one Higgs doublet, or with fully measured boundary values ($\widehat\alpha^{(5)}(M_Z^2)$, $\sin^2\widehat\theta_W=0.23129$, $\alpha_s=0.1180$, the familiar Standard-Model starting point), the three lines still fail to meet; the non-unification is an artefact neither of the doublet count nor of the asymptotic electromagnetic input.

The interpretation is the same as in version 3, and survives the revision of the weak sector: with an essentially Standard-Model $\beta$-function content, the $\EomE$ structure is an \emph{algebraic} unification of the pre-symmetry-breaking structure, not a running-coupling unification in the sense of Refs.~\cite{GeorgiGlashow1974,GeorgiQuinnWeinberg1974}.  A demonstration of high-scale running unification would require additional intermediate-scale matter or thresholds, which would be a different, nonminimal hypothesis, to be analysed together with the boundary construction.

\subsection{Infrared QCD running}
\label{sec:ir-running}

In the infrared the QCD $\beta$-function is standard, so evolving the derived $\alpha_3^{(0)}=\alphastar$ downward traces the usual running curve displaced by the boundary offset.  Using three-loop $\MSbar$ running~\cite{LarinVermaseren1993} with leading-order continuous matching at $m_b(m_b)=4.18$ GeV, the framework curve sits $1.06\%$ below the PDG-normalised curve at $M_Z$, $-2.08\%$ at $m_b$, and $-3.06\%$ at $m_\tau=1.777$ GeV (\cref{tab:alphas-ir}).  The displayed interval crosses only the bottom threshold: its lowest point, $m_\tau$, is above $m_c(m_c)=1.27$ GeV.  Finite higher-order decoupling corrections~\cite{Chetyrkin1998} are omitted in this diagnostic; the prescription and the code branch for continuation below $m_c$ are specified in \cref{app:running-repro}.  The deficit grows into the infrared because $\alpha_s$ is larger and the running steeper there.  This consistency check is useful---the boundary value could have been grossly wrong---but it is not independent confirmation: the curve being reproduced is the Standard-Model curve, and the framework enters only through the single boundary number, whose required correction is already recorded in \cref{eq:Z3}.

\begin{table}[t]
\centering
\caption{Infrared running of the strong coupling.  ``Framework'' starts from $\alpha_3^{(0)}=0.11675418$; ``PDG-normalised run'' starts from $\alpha_s(M_Z)=0.1180$.  Both use three-loop $\MSbar$ running with leading-order continuous matching at the bottom threshold only; the row at $m_b$ is the $n_f=5$ value and the lower rows use $n_f=4$.  The displayed interval stops above the charm threshold.  Finite higher-order decoupling corrections are not included.}
\label{tab:alphas-ir}
\begin{tabular}{lccc}
\toprule
Scale $\mu$ & $\alpha_s^{\rm fw}(\mu)$ & $\alpha_s^{\rm PDG\ run}(\mu)$ & deficit \\
\midrule
$M_Z$              & 0.11675 & 0.11800 & $-1.06\%$ \\
$10~\mathrm{GeV}$  & 0.17525 & 0.17815 & $-1.62\%$ \\
$m_b(m_b)=4.18~\mathrm{GeV}$ & 0.21968 & 0.22434 & $-2.08\%$ \\
$3~\mathrm{GeV}$   & 0.24663 & 0.25261 & $-2.37\%$ \\
$2~\mathrm{GeV}$   & 0.29118 & 0.29977 & $-2.87\%$ \\
$m_\tau=1.777~\mathrm{GeV}$ & 0.30767 & 0.31739 & $-3.06\%$ \\
\bottomrule
\end{tabular}
\end{table}

\section{The K\'arolyh\'azy correction, with its identifiability made explicit}
\label{sec:karolyhazy}

The 2022 paper also considered a K\'arolyh\'azy-type length uncertainty~\cite{Karolyhazy1966,Singh2022EPJP} as an infrared correction capable of closing the residual electromagnetic gap.  A coefficient must be retained:
\begin{equation}
 (\Delta\ell)^3=c_K L_P^2\ell.
 \label{eq:Krelation}
\end{equation}
With $\ell=1/\mu$ and $L_P=1/M_P$ in natural units,
\begin{equation}
 \delta_K:=\frac{L_P}{\Delta\ell}
 =c_K^{-1/3}\left(\frac{\mu}{M_P}\right)^{1/3}.
 \label{eq:deltaK}
\end{equation}
The corrected electromagnetic expression is
\begin{equation}
 \alpha_\gamma=A^2\left(\sqrt{\frac{3}{32}}+\delta_K\right)^4.
 \label{eq:alphaK}
\end{equation}
Matching the CODATA 2022 value~\cite{CODATA2022} requires
\begin{equation}
 \delta_K=2.26865496\times10^{-6},
 \qquad
 \frac{\mu}{c_K}=M_P\delta_K^3=142.55\ \mathrm{GeV}.
 \label{eq:Kresult}
\end{equation}
Thus the electromagnetic datum determines only the combination $\mu/c_K$.  More importantly, \cref{eq:alphaK} has been solved using the same measured $\alpha(0)$ that defines $Z_\gamma$ in \cref{eq:Zgamma}; it is a reparametrisation of the required correction, not an independent calculation of $Z_\gamma$.  Setting $c_K=1$ gives an electroweak-scale number, but it is not a prediction unless the coefficient, the localisation scale and the way the uncertainty enters the photon kinetic term are derived independently of $\alpha(0)$.  The inferred scale is $\mu/c_K\simeq143$ GeV (an inverse length), not a length of order $143\ \mathrm{GeV}^{-1}$.

\section{Phenomenological comparison}
\label{sec:phenom}

\Cref{tab:comparison} gives an even-handed numerical summary.  Experimental errors are quoted only to show the scale of present tension; no theoretical uncertainty has been assigned to the conditional formulas.

\begin{table}[ht]
\centering
\small
\caption{Comparison with the PDG 2024/2025 and CODATA 2022 reference values~\cite{PDG2024,CODATA2022}.  The fine-structure uncertainty is metrological, so a small fractional difference is not precision agreement.  The weak-angle comparison uses the running $\MSbar$ quantity appropriate to the coupling ratio; as a low-$Q^2$ reference, the SLAC E158 effective angle is $0.2397\pm0.0013$~\cite{E158}.}
\label{tab:comparison}
\begin{tabularx}{\textwidth}{@{}>{\raggedright\arraybackslash}p{0.27\textwidth}cc>{\raggedright\arraybackslash}X@{}}
\toprule
quantity & conditional value & experiment & assessment \\
\midrule
$\alpha_s(M_Z)$ & $0.11675418$ & $0.1180\pm0.0009$ & $-1.06\%$; $-1.4\sigma$; colour matching conditional \\
$\alpha(0)$ & $0.00729713629$ & $0.0072973525643(11)$ & $-2.96\times10^{-5}$ fractionally; not metrological agreement \\
$\sin^2\widehat\theta_W$ (equal-connection benchmark) & $0.23076923$ & $0.23129\pm0.00004$ & $-0.225\%$; $-13\sigma$ if exact; not embedding independent \\
$\sin^2\widehat\theta_W$ (old, withdrawn) & $\simeq0.2497$ & $0.23129\pm0.00004$ & $\simeq+8.0\%$; formal $\simeq+460\sigma$; not normalisation invariant \\
$\alpha_s(M_Z)/\widehat\alpha(M_Z)$ & $16$ & $15.096\pm0.115$ & $+5.99\%$; same-scale benchmark misses by $7.9\sigma$ \\
$\alpha_s(M_Z)/\alpha(0)$ & $16$ & $16.170\pm0.123$ & $-1.05\%$; mixed-regime comparison, $-1.4\sigma$ \\
\bottomrule
\end{tabularx}
\end{table}
\FloatBarrier

The scientifically relevant improvement is not that $3/13$ sits near the measured weak angle.  It is that the old $8\%$ discrepancy and field-normalisation ambiguity have been replaced by a conventional kinetic-and-current problem.  The audit shows both an internally consistent equal-connection benchmark and a trace-induced alternative that does not yield the same factor.  The projected localisation/interface action must select the kinetic metric before a weak-angle prediction exists.

\section{What must be derived next}
\label{sec:open}

The following calculations would decide whether the present conditional programme can become a first-principles result.  The ordering is deliberate: quantities must be derived before comparison with $\alpha(0)$.

\begin{enumerate}[label=\textbf{O\arabic*.}]
\item \textbf{Derive the spectral coefficient from the action.}  Building on the structural spectral-action map of Ref.~\cite{SinghSpectral2026}, the coefficient denoted $\alphastar$ here must be obtained in the form $a_\star=I_2\chi(q_0)$, or its correct replacement, from the projected trace-dynamics action without consulting $\alpha(0)$.  This calculation must derive the multiplicative character, the lower-root choice and the surviving power of the Jordan invariant rather than selecting them because of their numerical consequence.

\item \textbf{Project the photon kinetic norm.}  Compute the projected second variation $H_{AB}(X_0)=\delta^2S_{\rm loc}/(\delta X_A\delta X_B)|_{X_0}$, resolve the physical blocks, remove gauge and ghost directions, and project the photon zero mode independently.  Only the resulting $K_\gamma$ can decide whether the Hilbert--Schmidt normalisation used in \cref{eq:Fgamma} is physical and whether the $1/6$ survives canonical normalisation.

\item \textbf{Derive the photon localisation spectrum and $Z_\gamma$.}  Calculate the normalised photon zero mode, a positive physical transverse spectrum, the charged chiral Dirac zero mode, and all charged excited-state masses and overlaps in a declared four-dimensional matching scheme.  The target inferred from data is $Z_\gamma=1.0000296379\ldots$, or $K_\gamma^{\rm phys}/K_\gamma^{(0)}=0.99997036296\ldots$, but neither number may be used as input.

\item \textbf{Respect the negative localisation tests.}  A detailed companion analysis~\cite{SinghLocalization2026} tests the canonical $A_2\times E_6$ route.  Conditional on the channel-scalar quadratic block displayed in the current action, it does not presently provide a photon-specific sixfold trace or the required connected Hessian: compact reality pairs the six $A_2$ roots into three real sectors, the $(\bm8,\bm1)$ block is neutral under an $E_6$ photon, and the mixed $(\bm3,\bm{27})\oplus(\bar{\bm3},\overline{\bm{27}})$ multiplicity is universal over $E_6$ generators and risks counting matter already present.  A neutral six-link sector likewise gives no electromagnetic threshold by itself.  Any successful six-channel model must therefore derive both its incidence operator and equal kinetic weights from a sector-specific localisation mechanism.

\item \textbf{Derive the two-abelian kinetic/current embedding and check $SU(2)_L$.}  Starting from the $L_{BB}+L_{BF}+L_{FF}$ decomposition, identify the Stage-1 Cartan and Cl$(6)$ trace connections after localisation, derive the coefficients in \cref{eq:twoU1kin,eq:parentcharges}, and show which order parameter removes the orthogonal combination.  The same map must demonstrate whether the geometric/internal $SU(2)_L$ fusion produces one kinetic term or two.  The equal-connection value $\eta=1/2$ is one benchmark; the data-inferred value is $0.501468\pm0.000113$, but neither is presently derived.

\item \textbf{Derive colour matching from an adjoint-covariant profile family.}  The Schur argument applies to the colour-singlet photon.  It does not determine the eight gluon profiles.  The localisation map must construct their adjoint covariance, kinetic metric and coupling, thereby testing both $d_3=1$ and the restricted bound \cref{eq:cap8}.

\item \textbf{Complete six-dimensional and four-dimensional matching.}  The current six-dimensional interface analysis~\cite{SinghInterface2026}, audited in Ref.~\cite{SinghLocalization2026}, supplies a useful framework but no nonzero correction in the solved free diagnostics: the weak-scale electron example leaves only the elementary non-tachyonic transverse zero mode, and the solved Gaussian product model has $PHQ=0$.  These statements are not a no-go theorem for interacting six-dimensional localisation.  The missing photon transverse Hessian, physical charged spectrum and interaction vertices must be derived before a six-dimensional threshold can contribute to $Z_\gamma$.  Finite matching must then include the actual charged spectrum in one declared scheme.

\item \textbf{Derive the regime map.}  The present framework distinguishes asymptotic abelian and broken non-abelian readings, but a rule mapping these to $Q=0$, $M_Z$, or a localisation scale is open.  Until that rule exists, the same-scale ratio must remain the conservative benchmark.
\end{enumerate}

These are falsifiable tasks.  Failure to derive the photon-specific kinetic norm or the spectral coefficient independently of $\alpha(0)$ would leave the fine-structure proximity coincidental.  A microscopic two-$U(1)$ result selecting the trace-induced alternative \cref{eq:traceinduced}, or two identical $SU(2)_L$ kinetic terms, would invalidate the equal-connection weak benchmark.

\section{Conclusions}
\label{sec:conclusions}

The three gauge sectors can be placed in a cleaner conditional framework than in earlier versions of this paper.  Within the specified construction, the algebraic inputs are the primitive charge $q_0=1/3$, the exceptional-Jordan spacing $\delta^2=3/8$, and the one-generation trace identities.  The action parameter $\alpha_{\rm TD}$ is distinct from the electromagnetic fine-structure constant.  The earlier electromagnetic expression is reorganised into a spectral-character factor and a broken-phase support factor.  The real Schur lemma fixes the invariant photon-profile direction, but the physical $1/6$ requires the still-uncomputed projected kinetic norm.  Consequently $[\alpha_\gamma^{(0)}]^{-1}=137.04006064\ldots$ is a strong conditional boundary value, not a first-principles prediction, and the data-inferred $Z_\gamma$ remains to be calculated.

The previous weak-angle calculation remains withdrawn.  The revised two-$U(1)$ treatment now assigns charges $c_1Y$ and $c_2Y$ with $c_1+c_2=1$, displays a difference-charged link field, and proves that the massless diagonal connection couples to exactly $Y$.  Equal connection coefficients give $g_Y^2/g_2^2=3/10$ and $\sin^2\theta_W=3/13$, but applying trace normalisation to the scaled generators gives a different result.  The number $3/13$ is therefore an illustrative benchmark, not an embedding-independent prediction.  Treated as exact, it is about $13\sigma$ from the quoted running value; model incompleteness is not used as a statistical error.

The resulting coupling ledger,
\[
 d_3^2=1,
 \qquad d_2^2=\frac{13}{48},
 \qquad d_Y^2=\frac{13}{160},
 \qquad d_\gamma^2=\frac1{16},
\]
is compact and internally consistent only under the stated equal-connection hypothesis.  Four limitations remain.  First, the abelian $Q=0$ versus colour $M_Z$ comparison uses a phenomenological regime selection; the conservative same-scale ratio misses $16$ by $6\%$.  Second, the photon kinetic norm, the overlap rule and $d_3=1$ are underived; the colour bound at $8$ applies only to individually $J$-invariant profiles, not to a general adjoint-covariant gluon family.  Third, the companion audit~\cite{SinghLocalization2026} shows that the tested canonical $A_2$ descent and solved six-dimensional free-product models do not generate a photon-specific sixfold norm or a nonzero $Z_\gamma$; the Hessian conclusion is conditional on the channel-scalar quadratic block currently displayed.  Fourth, the illustrative boundary values do not unify under minimal-desert running.  The paper should accordingly be read as a normalisation audit and a falsifiable conditional gauge-sector programme, not as a parameter-free prediction of Standard-Model couplings.

\appendix

\section{Trace table and anomaly arithmetic}
\label{app:traces}

\begin{table}[ht]
\centering
\caption{One Standard-Model generation in the convention $Q=T_L^3+Y$.  Multiplicity includes colour and weak components.}
\label{tab:charges}
\begin{tabular}{@{}lcccc@{}}
\toprule
field & multiplicity & $Q$ & $T_L^3$ & $Y$ \\
\midrule
$u_L$ & $3$ & $2/3$ & $1/2$ & $1/6$ \\
$d_L$ & $3$ & $-1/3$ & $-1/2$ & $1/6$ \\
$u_R$ & $3$ & $2/3$ & $0$ & $2/3$ \\
$d_R$ & $3$ & $-1/3$ & $0$ & $-1/3$ \\
$\nu_L$ & $1$ & $0$ & $1/2$ & $-1/2$ \\
$e_L$ & $1$ & $-1$ & $-1/2$ & $-1/2$ \\
$\nu_R$ & $1$ & $0$ & $0$ & $0$ \\
$e_R$ & $1$ & $-1$ & $0$ & $-1$ \\
\bottomrule
\end{tabular}
\end{table}

From \cref{tab:charges},
\begin{align}
 \Tr Q^2
 &=3\left[\frac49+\frac19+\frac49+\frac19\right]+1+1
 =\frac{16}{3},\\
 \Tr(T_L^3)^2
 &=3\left[\frac14+\frac14\right]+\frac14+\frac14=2,\\
 \Tr Y^2
 &=3\left[2\cdot\frac1{36}+\frac49+\frac19\right]+2\cdot\frac14+1
 =\frac{10}{3},\\
 \Tr(YT_L^3)
 &=3\left[\frac16\cdot\frac12+\frac16\left(-\frac12\right)\right]
 +\left(-\frac12\right)\frac12+\left(-\frac12\right)\left(-\frac12\right)=0.
\end{align}
For anomaly sums, right-handed fields are converted to left-handed conjugates, reversing the abelian charge.  The explicit sums in \cref{eq:a33Y,eq:a22Y,eq:agravY,eq:aYYY} then follow.

\section{Matrix form of the current-consistent \texorpdfstring{$U(1)\times U(1)$}{U(1) x U(1)} reduction}
\label{app:diagonal}

Let $A=(A^{(1)},A^{(2)})^T$, $n=(1,1)^T$, $r=(1,-1)^T$ and $c=(c_1,c_2)^T$.  The quadratic gauge and visible-current terms, including abelian kinetic mixing~\cite{Holdom1986} and diagonal product-group breaking~\cite{ArkaniHamed2001}, can be written
\begin{equation}
 \mathcal L=-\frac14F^T\mathcal K F+J_Y^\mu c^TA_\mu
 +\frac12m_\ell^2(r^TA_\mu)(r^TA^\mu),
 \qquad c^Tn=1,
 \label{eq:matrixU1}
\end{equation}
where $\mathcal K$ is positive definite.  The link mass imposes $A=nA_Y$ at energies below $m_\ell$.  Therefore
\begin{equation}
 \mathcal L_{\rm eff}=-\frac14(n^T\mathcal K n)F_Y^2+J_Y^\mu A_{Y\mu},
 \qquad \frac1{g_Y^2}=n^T\mathcal K n.
 \label{eq:matrixeff}
\end{equation}
The canonically normalised field is $B_Y=A_Y/g_Y$.  A massive canonical direction may be chosen $\mathcal K$-orthogonal to $n$; it need not be the Euclidean difference vector when kinetic mixing is present.

For fixed equal connection coefficients,
\begin{equation}
 \mathcal K_{\rm conn}=\frac1{g_{Y,0}^2}
 \begin{pmatrix}1&\chi\\ \chi&1\end{pmatrix},
 \qquad \eta_{\rm conn}=\frac1{2(1+\chi)}.
 \label{eq:Kconn}
\end{equation}
For trace-induced normalisation of $q_i=c_iY$,
\begin{equation}
 \mathcal K_{\rm tr}=\frac1{g_{Y,0}^2}
 \begin{pmatrix}c_1^2&\rho c_1c_2\\ \rho c_1c_2&c_2^2\end{pmatrix},
 \qquad \eta_{\rm tr}=\frac1{c_1^2+c_2^2+2\rho c_1c_2}.
 \label{eq:Ktrace}
\end{equation}
Equations~\eqref{eq:Kconn} and \eqref{eq:Ktrace} are not related by varying the matter weights while keeping the kinetic coefficients fixed; they are distinct microscopic hypotheses.  This is the normalisation ambiguity that the projected action must resolve.

\section{Sensitivity to the support count and to the spectral datum}
\label{app:sensitivity}

This sensitivity exercise concerns conditional algebraic boundary coefficients, before the running and matching factors $Z_i$ are applied.  It retains the linear-overlap rule and equal-trace parent normalisation of \cref{sec:strong-em}, assumes that the auxiliary support norm is the norm induced by the projected photon kinetic term so that the physical squared-coupling dilution is $1/N$, and imposes undiluted colour matching $d_3=1$.  Let $N$ denote the effective support dimension ($N=\dim H_6=6$ in the stated model, varied here only as a sensitivity parameter) and let $x$ replace the spectral datum $\delta^2=3/8$.  Only under these assumptions does the spectral-character ansatz give
\begin{equation}
 \frac{\alpha_s}{\alpha_{\rm em}}=\frac83\,N,
 \qquad
 \alpha_s(x)=x^2\exp\!\left[2q_0\!\left(q_0-\sqrt{x}\right)\right],
 \qquad
 \alpha_{\rm em}(x,N)=\frac{3}{8N}\,\alpha_s(x),
 \label{eq:sensitivity}
\end{equation}
with $q_0=1/3$; here $\alpha_s$ and $\alpha_{\rm em}$ abbreviate the boundary coefficients $\alpha_3^{(0)}$ and $\alpha_\gamma^{(0)}$.  The present construction corresponds to $(N,x)=(6,3/8)$.  The result is highly sensitive to the discrete support dimension: $N=5$ and $N=7$ would give $\alpha_s/\alpha_{\rm em}=40/3\approx13.3$ and $56/3\approx18.7$, both far from the observed mixed-regime ratio.  The dependence on $x$ is smoother but significant: near $x=3/8$,
\begin{equation}
 \frac{\delta\alpha_s}{\alpha_s}
 \approx\left(2-\frac{\sqrt{x}}{3}\right)\frac{\delta x}{x}
 \approx1.80\,\frac{\delta x}{x}.
\end{equation}
In the present formulation $N=6$ is the dimension of the proposed support space $H_6$, fixed by the three complex ladder directions rather than fitted as a continuous parameter.  This converts the numerical choice into a structural and falsifiable hypothesis, but not yet a microscopic derivation: the projected kinetic trace could pair the roots, count matter multiplicities, or yield a different physical norm.  The datum $x=3/8$ is fixed as the spacing of the adopted Jordan family matrices.  The remaining uncertainty lies in the map from those algebraic objects to the localised gauge kinetic term.

\section{Reproducibility of the numerical results}
\label{app:running-repro}

The numerical values in \cref{sec:matching} use $M_Z=91.1880$ GeV, $\alpha_s(M_Z)=0.1180$, $\widehat\alpha^{(5)}(M_Z^2)^{-1}=127.930$, and $\sin^2\widehat\theta_W(M_Z)=0.23129$~\cite{PDG2024}, and the 2022 CODATA fine-structure constant~\cite{CODATA2022}.  Electroweak evolution uses \cref{eq:EWrun}; the two-loop cross-check uses the Machacek--Vaughn gauge coefficients in GUT normalisation ($g_1^2=\tfrac53 g_Y^2$) with the top-Yukawa term at fixed $y_t(M_Z)=0.99$.  The near-electroweak QCD diagnostic in \cref{sec:running} uses \cref{eq:QCDrun} with $n_f=5$ and no threshold crossing.  The three-loop cross-check and infrared evolution add the standard $\MSbar$ coefficient~\cite{LarinVermaseren1993,PDG2024}, in the convention
\[
 \left.\frac{d\alpha_s}{d\ln\mu}\right|_{\text{three loops}}
 =-\frac{\beta_0}{2\pi}\alpha_s^2-\frac{\beta_1}{8\pi^2}\alpha_s^3
 -\frac{\beta_2}{32\pi^3}\alpha_s^4,
 \qquad
 \beta_2=\frac{2857}{2}-\frac{5033}{18}n_f+\frac{325}{54}n_f^2.
\]
For \cref{tab:alphas-ir}, the code switches from $n_f=5$ to $n_f=4$ at $m_b(m_b)=4.18$ GeV, imposing continuity of $\alpha_s$.  Since $m_\tau=1.777$ GeV exceeds $m_c(m_c)=1.27$ GeV, the displayed table does not cross the charm threshold.  The code also implements continuous $n_f=4$ to $n_f=3$ matching for runs continued below $m_c$.  This is a leading-order matching prescription, not full higher-order decoupling: the finite threshold terms of Ref.~\cite{Chetyrkin1998} are not included.  At $\mu=m_h(\mu)$ the one-loop matching logarithm vanishes, but a finite two-loop decoupling term remains; consequently the quoted table is a three-loop-running diagnostic with truncated threshold matching.  The ultraviolet diagnostic uses \cref{eq:uv-bc} with $(b_1,b_2,b_3)=(21/5,-3,-7)$.  The accompanying source bundle contains the numerical scripts, their output, and the scripts that reproduce all four figures.  The scale exercises are intentionally minimal: they are diagnostics, not substitutes for a full threshold analysis with the actual beyond-Standard-Model spectrum of the $\EomE$ construction.

\label{end:running-repro}

\begin{samepage}
\section*{Author declarations}
\label{declarations}

\paragraph{Data availability.}
All numerical inputs are stated in the text.  Reproducibility scripts, numerical output and figure sources accompany the submission source bundle.  In this study, no new data were created or analysed.

\paragraph{Competing interests.}
The author declares no competing interests.

\paragraph{Use of generative AI.}
During the preparation of this manuscript, the author used OpenAI's GPT-5.6 Sol and Codex, and Anthropic's Claude Fable 5, in adversarial mode for support in technical checking, organisation, writing and editing.  The original ideas are due to the author.  The author takes full intellectual responsibility for the content of the manuscript.

\end{samepage}

\end{document}